\documentclass[aps,prl,reprint,longbibliography,floatfix,superscriptaddress]{revtex4-2}
\usepackage{color}
\usepackage{amsmath}
\usepackage{amssymb}
\usepackage{amsfonts}
\usepackage{gensymb}
\usepackage{graphicx}
\usepackage{esint} 
\usepackage{ulem} 
\PassOptionsToPackage{normalem}{ulem}
\makeatletter
\usepackage{bm}
\usepackage{float}
\usepackage{xcolor}
\usepackage{xspace}
\usepackage{siunitx} 
\usepackage[version=4]{mhchem} 

\makeatother

\begin{document}
\title{Quasiparticle spectroscopy in tantalum films with different Ta/sapphire interfaces}

\author{Bicky Singh Moirangthem}
\affiliation{Ames National Laboratory, Ames, IA 50011, U.S.A.}
\affiliation{Department of Physics and Astronomy, Iowa State University, Ames, IA 50011, U.S.A.}

\author{Kamal R. Joshi}
\affiliation{Ames National Laboratory, Ames, IA 50011, U.S.A.}

\author{Anthony P. McFadden}
\affiliation{National Institute of Standards and Technology, Boulder, Colorado 80305, U.S.A.}

\author{Jin-Su Oh}
\affiliation{Department of Materials Science and Engineering, University of Virginia, Virginia, 22904, United States}
\affiliation{Ames National Laboratory, Ames, IA 50011, U.S.A.}

\author{Amlan Datta}
\affiliation{Ames National Laboratory, Ames, IA 50011, U.S.A.}
\affiliation{Department of Physics and Astronomy, Iowa State University, Ames, IA 50011, U.S.A.}

\author{Makariy A. Tanatar}
\affiliation{Ames National Laboratory, Ames, IA 50011, U.S.A.}
\affiliation{Department of Physics and Astronomy, Iowa State University, Ames, IA 50011, U.S.A.}

\author{Florent Lecocq}
\affiliation{National Institute of Standards and Technology, Boulder, Colorado 80305, U.S.A.}

\author{Raymond W. Simmonds}
\affiliation{National Institute of Standards and Technology, Boulder, Colorado 80305, U.S.A.}

\author{Lin Zhou}
\affiliation{Department of Materials Science and Engineering, University of Virginia, Virginia, 22904, United States}
\affiliation{Ames National Laboratory, Ames, IA 50011, U.S.A.}

\author{Matthew J. Kramer}
\affiliation{Ames National Laboratory, Ames, IA 50011, U.S.A.}

\author{Ruslan Prozorov}
\email{Corresponding author: prozorov@ameslab.gov}
\affiliation{Ames National Laboratory, Ames, IA 50011, U.S.A.}
\affiliation{Department of Physics and Astronomy, Iowa State University, Ames, IA 50011, U.S.A.}

\date{6 March 2026}

\begin{abstract}
One of the crucial aspects of current research in quantum information science is the identification and control of loss mechanisms in superconducting circuits. Although microwave measurements directly quantify device performance, additional techniques that probe quasiparticle excitations in superconducting films are needed to understand the microscopic mechanisms underlying dissipation and decoherence. Here, we present results from quasiparticle spectroscopy of Ta/sapphire films by measuring the Meissner-state magnetic susceptibility using a precision frequency-domain resonator specifically designed for thin films. We find direct evidence for additional low-energy excitations in samples with lower internal quality factors. These excitations are consistent with deep subgap states due to two-level systems, Yu-Shiba-Rusinov states near the gap edge, and perhaps other pair-breaking mechanisms. The developed non-destructive frequency-domain quasiparticle spectroscopy is a valuable addition to the quantum materials toolbox.
\end{abstract}
\maketitle

\section{Introduction}

Superconducting (SC) qubits are leading candidates for implementing large-scale quantum computers because they exploit well-defined macroscopic quantum coherence and are compatible with scalable circuit architectures \cite{Reagor2016,Kjaergaard2020,Leon2021}. In state-of-the-art devices, energy relaxation and dephasing are often limited by extrinsic microscopic dissipation channels that are not present in an ideal superconducting condensate \cite{Siddiqi2021}. Pervasive defects give rise to two-level systems (TLS) and magnetic moments associated with vacancies and impurities, commonly found in structurally disordered or amorphous oxides and interfacial layers \cite{Siddiqi2021,Burnett2016,Muller2019,Gao2008,Oh2024}. These defects can couple to microwave fields, generating dielectric loss, and their influence is amplified in planar architectures where electric fields concentrate near surfaces and interfaces \cite{Martinis2005}. Consequently, understanding how processing and microstructure affect interfacial properties remains crucial for advancing coherence times and reproducibility.

Tantalum has recently emerged as an attractive materials platform for SC transmon-style qubits and related circuits because, in some experiments, Ta films exhibit exceptionally low microwave loss and longer lifetimes \cite{Ganjam2024,Wang2022,Place2021}. At the same time, device performance is sensitive to growth conditions, phase purity, and, most importantly, the chemical and structural state of the Ta/oxide and Ta/substrate interfaces \cite{Ganjam2024,Crowley2023}. Elemental Ta is a conventional, fully gapped $s$-wave superconductor; thus, any measurable low-energy excitations at $T \ll T_c$ suggest additional physics beyond a clean BCS limit \cite{Bardeen1957}. Such excitations reflect the existence of subgap states from pair-breaking due to anisotropy, defects in crystalline lattice, bound atomic complexes, magnetic moments, proximity effects at interfaces, or other sources of quasiparticle excitations \cite{Henderson1972,Koch1974,Dynes1978,Phillips1987,Graaf2020,Gurevich2017,Bafia2021,Ghimire2024,abogoda2025a,he2025a,ueki2025,abogoda2025,zarea2025}. 

Their microscopic origin is the central question in current QIS research. Correlating superconducting-state spectroscopy with resonator performance can help narrow down plausible mechanisms.

These considerations motivate superconducting-state characterization methods that are directly sensitive to low-energy excitations. Although room-temperature structural and chemical probes provide indispensable information about microstructure and composition, they cannot determine the quasiparticle spectrum or the structure of the superconducting energy gap, which may contain additional low-energy states. Subgap features in the density of states — whether produced by disorder, impurities, or interfacial phenomena- manifest as non-activated (e.g., power-law) and often result in additional features, such as convex downturn upon cooling, bumps, and even non-monotonic temperature dependence of thermodynamic quantities such as the London penetration depth, $\lambda(T)$. Precision measurements of $\lambda(T)$ therefore provide an effective ``spectroscopic" window into the low-energy excitation spectrum of superconductors, complementing local probes and transport measurements \cite{Ghimire2024}.

In a recent study, McFadden \emph{et al.} reported a systematic study of the performance of the Ta devices, focusing on the Ta/sapphire interface \cite{lwn1-fznb}. They observed a nontrivial evolution of internal quality factors, $Q_i$, near the onset of epitaxial growth and, importantly, demonstrated that inserting a thin ($\sim$5 nm) Nb interlayer between sapphire and Ta can substantially reduce microwave loss for films deposited at the same temperature, 630$\degree$C. 

In this work, we study the electrodynamic response of the superconducting state of the same Ta films using a highly sensitive tunnel-diode resonator (TDR). We demonstrate that samples with the highest $Q_i$ exhibit a clean activated low-$T$ response consistent with a fully gapped spectrum, whereas lower-$Q_i$ films display distinct signatures of additional low-energy excitations that correlate with device performance.

\section{Experimental}

\subsection{Samples}

Three different types of films, with two samples each, were deposited at \SI{630}{\celsius} using different substrate preparations:
\begin{itemize}
\item \uline{Sample A:} 100 nm Ta deposited on a $c$-plane sapphire.
\item \uline{Sample B:} 95 nm Ta film deposited on a 5 nm thick Nb layer sputtered on a $c$-plane sapphire.
\item \uline{Sample C:} $c$-plane sapphire was treated with argon plasma at room temperature, followed by in situ deposition of 100 nm of Ta.
\end{itemize}

\begin{figure}[tb]
\includegraphics[width=8.5cm]{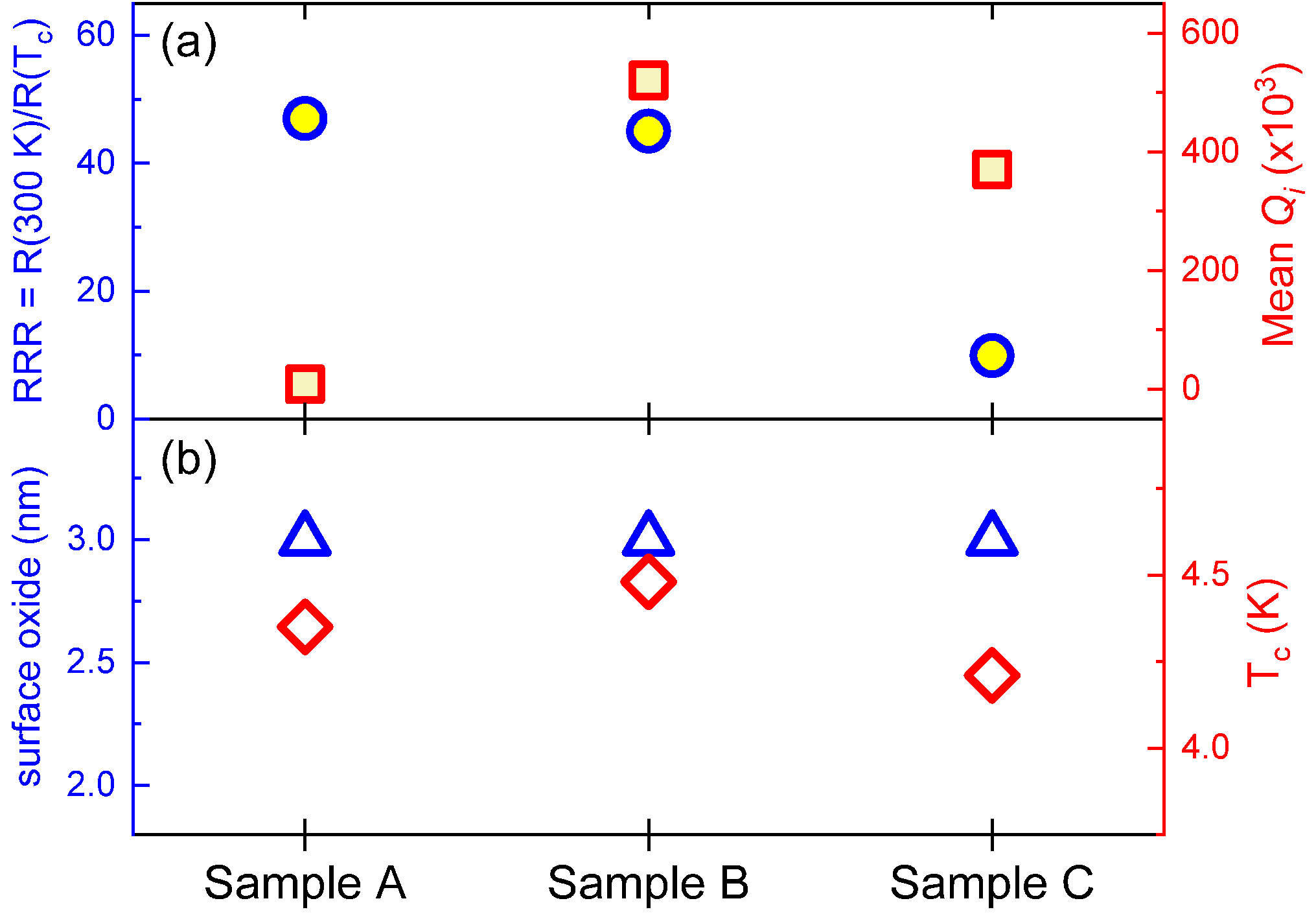}
\caption{(a) RRR and $Q_{i}$ for three different types of samples. (b) Corresponding surface oxide thickness and $T_{c}$. The data are taken from Table II of Ref.~\cite{lwn1-fznb}.}
\label{fig:sample_info}
\end{figure}

Detailed characterization of these films is reported elsewhere \cite{lwn1-fznb}. The key parameters, the residual resistivity ratio (RRR) and the average internal quality factor $Q_{i}$, measured in the single-photon limit, are shown in Fig.~\ref{fig:sample_info}(a). The thickness of the oxide layer at the Ta/air interface and $T_c$ are shown in Fig.~\ref{fig:sample_info}(b). Furthermore, substrate differences also affect the in-plane morphology (granular structure) of the studied films, which certainly influence, if not determine, transport properties. Not surprisingly, there is no obvious correlation between RRR, $Q_{i}$ and $T_{c}$ in all three sample types.

\subsection{ Electron microscopy }

Transmission electron microscopy (TEM) samples were prepared using a focused ion beam instrument (Thermo Fisher Scientific, Helios G3). The TEM samples were investigated using aberration-corrected TEM (Thermo Fisher Scientific, Titan Themis cubed) at 200 kV. A high-angle annular dark-field (HAADF) detector was used for dark-field imaging in STEM mode with a convergent semi-angle and a collection semi-angle of 18 mrad and 74-200 mrad, respectively.

\subsection{Quasiparticle spectroscopy}

The temperature-dependent variation of magnetic susceptibility $\chi$ in the Meissner state was measured using a self-oscillating frequency-domain tunnel-diode resonator (TDR) operating around \SI{10}{\mega\hertz}. This technique has been successfully used to study the energy-gap structure in various superconductors, and its sensitivity makes it well suited for thin films \cite{Van1975,giannetta2022london,Prozorov2006,Prozorov2011,Ghimire2024}.

In brief, the superfluid density, normalized by its value in the clean case, is:
\begin{equation}
n_{s}=1+\intop_{-\infty}^{\infty}\frac{\partial f\left(E\right)}{\partial E}\frac{N\left(E\right)}{N_{n}}\,dE
\label{eq:ns}
\end{equation}

\noindent where $E=0$ corresponds to the Fermi level, $N(E)$ is the quasiparticle density of states, $N_n$ is its normal-state value, and $f(E)$ is the Fermi function. The temperature dependence enters primarily through the derivative of the Fermi function $\partial f(E)/\partial E = -\mathrm{sech}^{2}(E/2t)/(4t)$,
where $t=T/T_{c}$ and $T_{c}$ is the measured superconducting transition temperature. At $T \to 0$, the derivative approaches a delta function. At finite temperatures, it broadens into a bell-shaped kernel that effectively sweeps through $N(E)$, thus revealing any significant subgap features. Since the London penetration depth is directly related to the superfluid density via $\rho_s(T)\equiv n_s(T)/n_s(0)=\left[\lambda(0)/\lambda(T)\right]^2$, measurements of $\chi(T)$, which is proportional to $\lambda(T)$ provide direct access to the quasiparticle spectrum. This is why we refer to this approach as ``gap spectroscopy.''

\section{Results}


\begin{figure*}[tbh]
\includegraphics[width=1\linewidth]{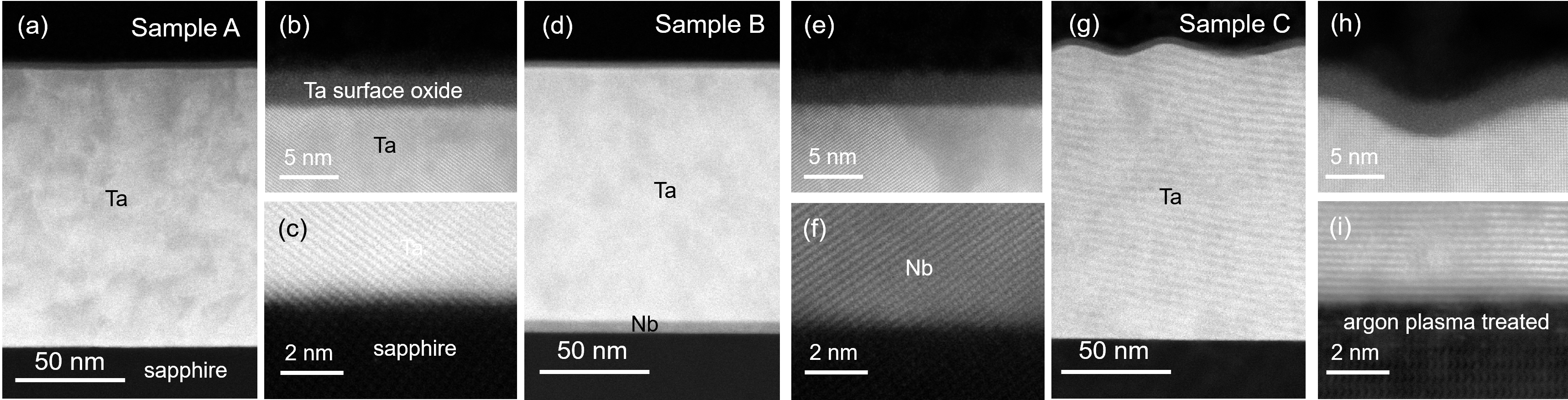} \caption{Cross-sectional scanning transmission electron microscopy (STEM) images of Samples A(a-c), B(d-f), and C(g-i). (a) Low mag HAADF-STEM images showing overall thin film on substrate structure and (b-c) atomic resolution HAADF-STEM images of surface and interface regions of Sample A. (d) Low mag HAADF-STEM images showing overall thin film on substrate structure and (e-f) atomic resolution HAADF-STEM images of surface and interface regions of Sample B. (g) Low mag HAADF-STEM images showing overall thin film on substrate structure and (h-i) atomic resolution HAADF-STEM images of surface and interface regions of Sample C.}
\label{fig:TEM}
\end{figure*}

Cross-sectional TEM was used to examine the overall microstructure, specifically the morphology, surface oxide layer, and metal-substrate interface of Samples A, B, and C. Figure~\ref{fig:TEM} shows HAADF-STEM micrographs of these three samples. While sample C displays a rough top surface, both Samples A and B exhibit flat surfaces (Fig.~\ref{fig:TEM}(b), (e), and (h)). Samples A and B show a well-defined metal-substrate interface, whereas sample C shows an irregular interface. Ar-plasma treatment of Sample C prior to deposition results in a damaged surface on the sapphire, which led to a different growth direction of the metal thin film compared to Samples A and B (Fig.~\ref{fig:TEM}(i)). This is also reflected in the in-plane granular structure of the studied films. Detailed orientation relationship between the metal morphology and the substrate in our samples is reported in Ref.\cite{lwn1-fznb}.

\begin{figure}[tb]
\includegraphics[width=8.5cm]{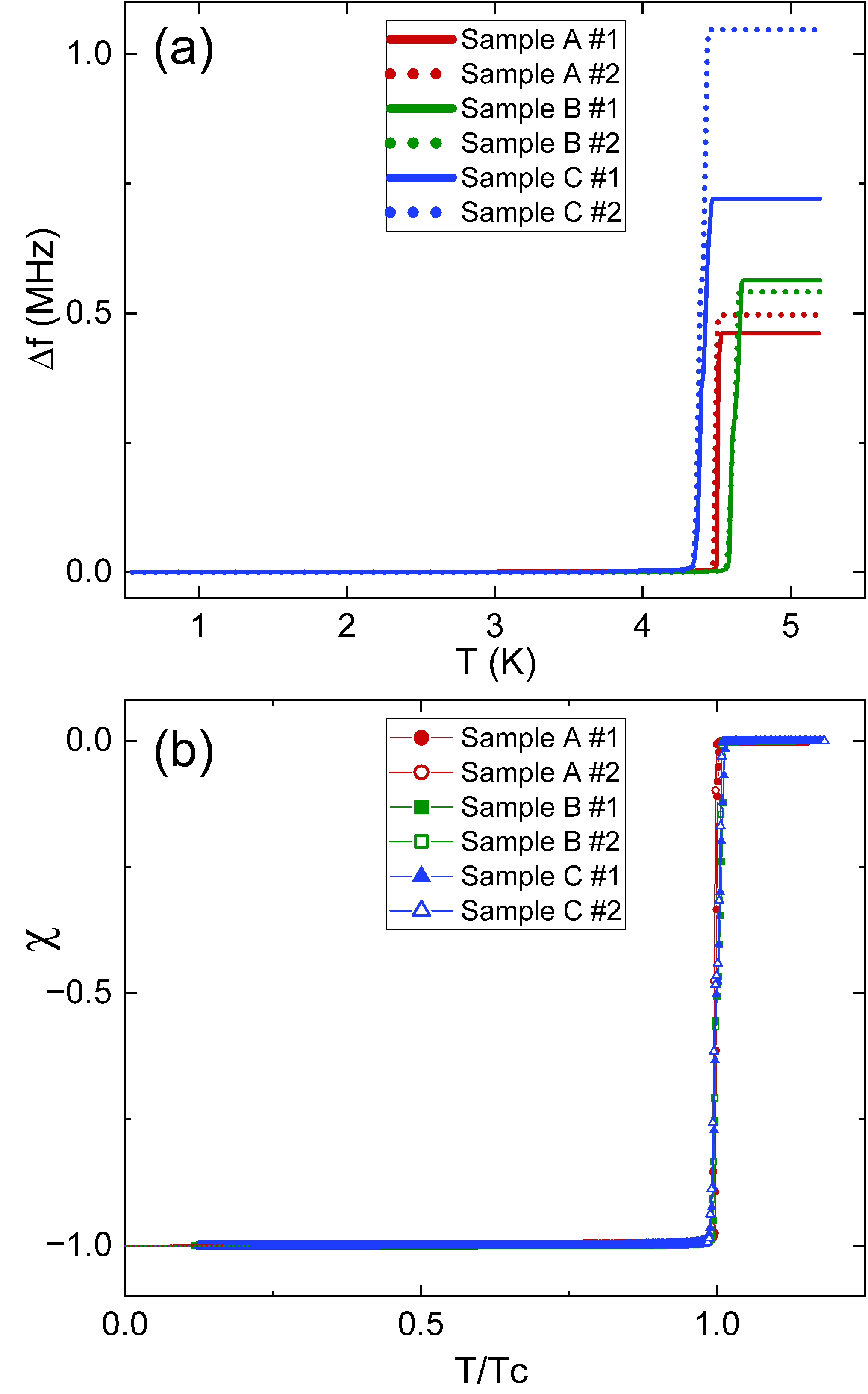}
\caption{(a) Change in the resonant frequency over the full temperature range in two tantalum samples each of A, B and C types. Superconducting transition temperatures are similar and consistent with transport measurements, where sample B has the highest and sample C has the lowest $T_{c}$. (b) Magnetic susceptibility obtained from the data in panel (a) as a function of normalized temperature, $t=T/T_{c}$.}
\label{fig:chi}
\end{figure}

Figure~\ref{fig:chi} shows the TDR-measured frequency change with temperature in six samples, two of each of the three sample types, labeled $\#1$ and $\#2$. The observed transition temperatures are consistent with the results from the transport measurements in Fig.~\ref{fig:sample_info}(b). Since the demagnetizing factors of such thin films are close to 1 when the field is applied perpendicular to the film \cite{prozorov2021meissner,Prozorov2000}, the frequency changes at low temperatures are very small; thus, an exceptionally high resolution provided by the TDR susceptometer is required. 

\begin{figure}[tb]
\includegraphics[width=8.5cm]{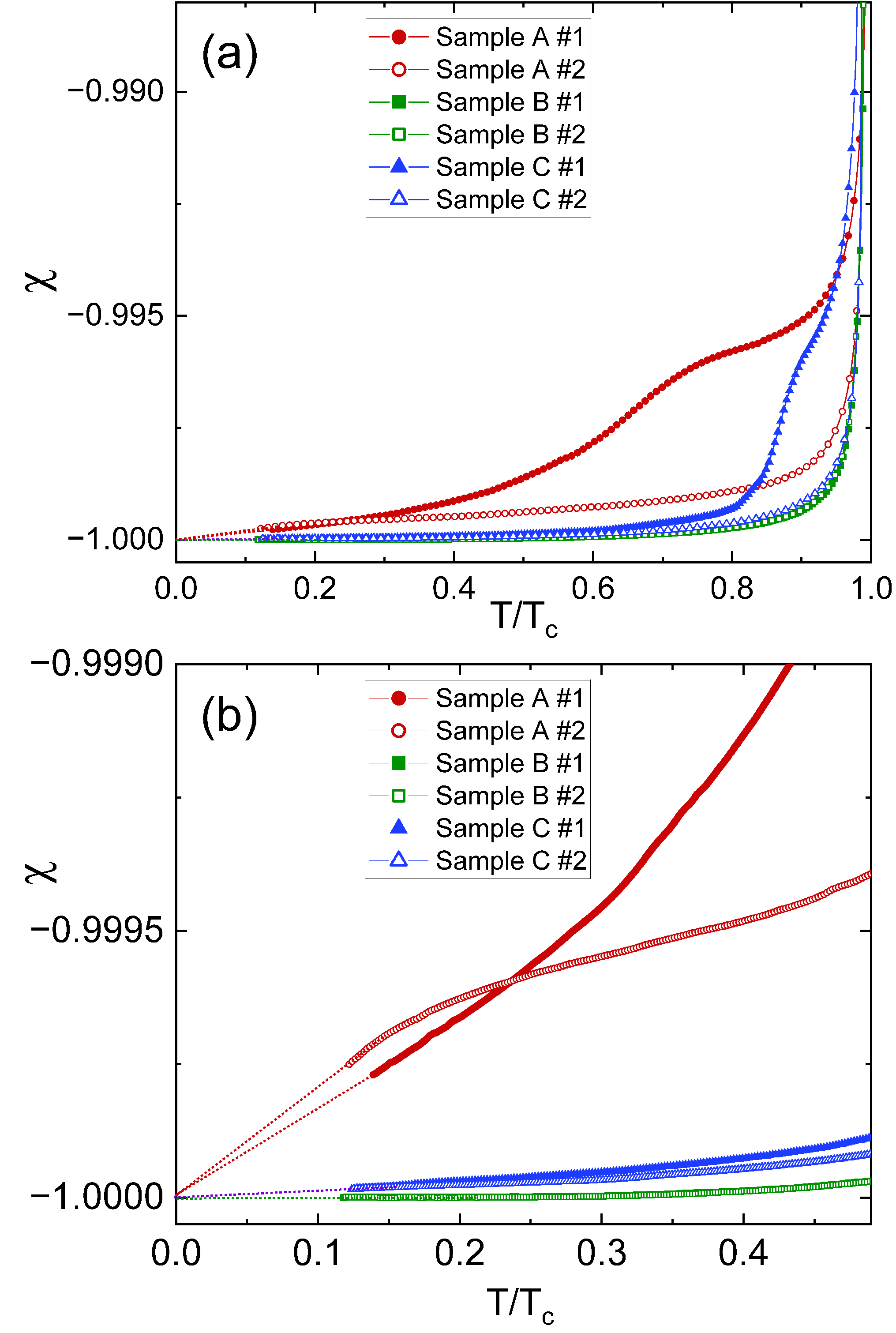}
\caption{(a) Low-temperature magnetic susceptibility in a restricted vertical scale highlighting deviations from a clean activated response. (b) Further zoom emphasizing the low-$T$ behavior used for power-law fitting. Compare the vertical scale with Fig.~\ref{fig:chi}.}
\label{fig:LT_chi}
\end{figure}

We now focus on lower temperatures. Figure~\ref{fig:LT_chi}(a) shows the interval up to $t=1$ on a restricted vertical scale from $-1$ to $-0.987$, and Fig.~\ref{fig:LT_chi}(b) zooms further to $t=0.5$ on a minuscule vertical scale from $-1$ to $-0.999$. Samples A $\#1$ and A $\#2$ (red symbols) clearly differ from B (green) and C (blue). Sample A $\#1$ exhibits a pronounced bump-like feature around $t\approx 0.7$, and sample A $\#2$ shows a convex downturn roughly below $t\approx 0.15$. Similarly, sample C $\#1$ shows a bump at a higher temperature around $t\approx 0.85$. In addition to bumps and downturns, the variation of $\chi(T)$ at low temperatures in samples of type A and type C is clearly non-exponential, which is unexpected for a clean $s$-wave superconductor. In stark contrast, a clean superconducting gap yields a smooth, monotonic temperature dependence that becomes exponentially attenuated upon cooling, as observed in the two B samples. To quantify the low-temperature variation, we fit the susceptibility to a power-law form $\chi(t)\propto t^{n}$. Exponential behavior is practically indistinguishable from a power-law with a large exponent of 4 and above over a low-temperature range of $t\leq 0.33$, so the large fitted values $n$ are consistent with activated behavior.

\begin{figure}[tbh]
\includegraphics[width=8.5cm]{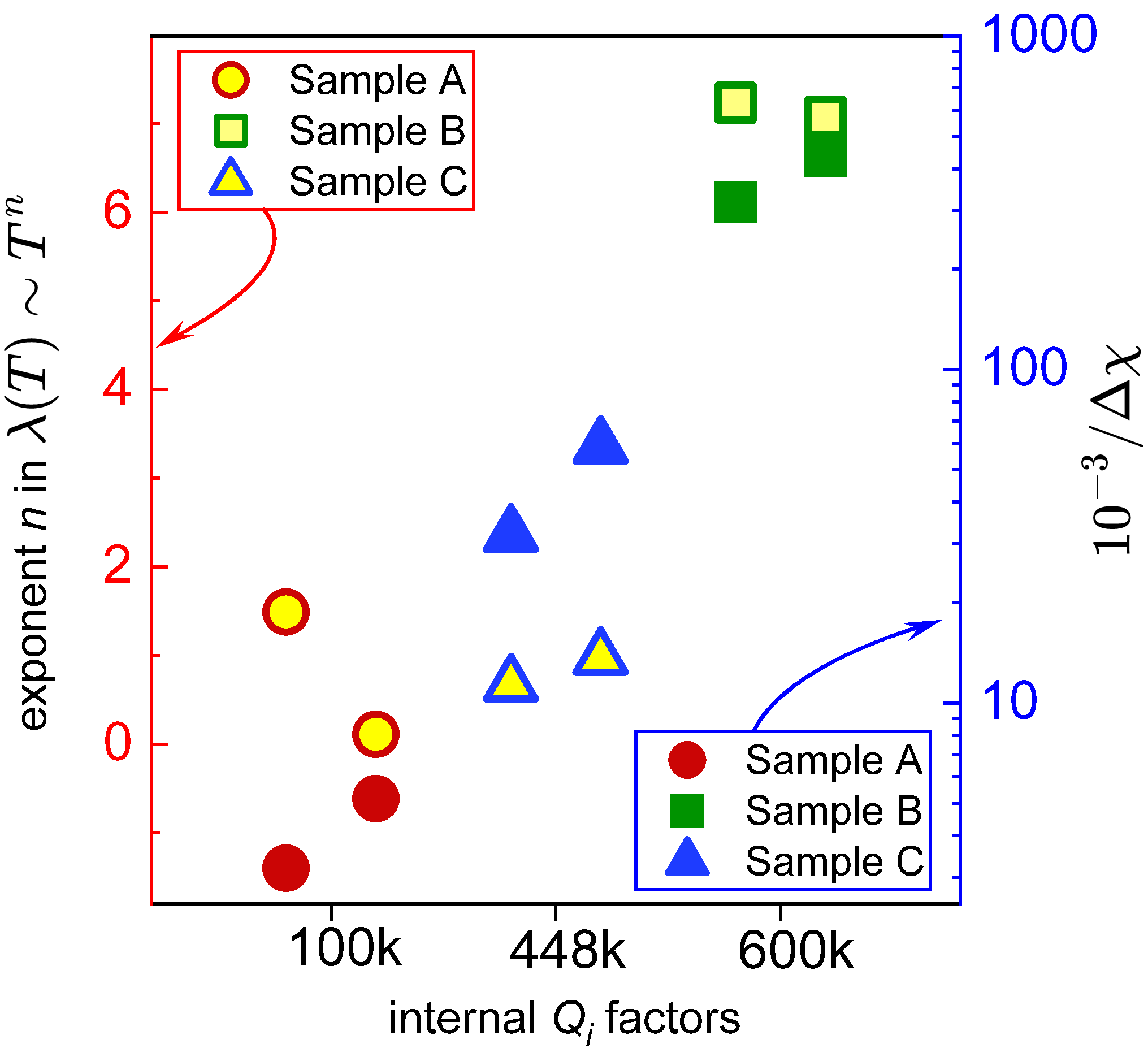}
\caption{Correlation between low-temperature susceptibility behavior and device performance. Left axis: power-law exponent $n$ from low-$T$ ($t\le0.33$) fits. Right axis: inverse susceptibility change $\Delta\chi^{-1}$ in the range $t\le 0.3$. The horizontal axis shows the internal quality factor $Q_i$.}
\label{fig:exponent}
\end{figure}

The exponents $n$ obtained from the low-temperature fitting are summarized on the left axis (red) in Fig.~\ref{fig:exponent}. To characterize the amount of quasiparticles generated at low temperatures, we plot the inverse change of susceptibility, $\Delta\chi^{-1}$, in the temperature interval up to $t=0.3$, on the right axis (blue) in Fig.~\ref{fig:exponent}. The inverse metric is selected to produce a larger number for higher-$Q_i$ samples, corresponding to the trend of the exponent $n$. Figure~\ref{fig:exponent} reveals a statistically significant correlation with the quality factors shown on the $x$-axis. Samples B, which have a clean superconducting gap with no evidence of in-gap bound states, show a strongly activated low-$T$ variation of $\chi(T)$ with $n>6$ and $\Delta\chi$ roughly two orders of magnitude smaller than in the lower-$Q_i$ samples.

\section{Discussion}

The London penetration-depth data reveal a systematic connection between superconducting-state spectral cleanliness and the microwave performance of the corresponding Ta-based devices. Samples of type B (Ta grown on a thin Nb interlayer) display essentially featureless, monotonic low-temperature susceptibility and an exponentially activated temperature dependence. This behavior is consistent with a fully gapped $s$-wave superconductor with a clean density of states. In contrast, samples of type A (Ta directly deposited on sapphire) and, to a lesser extent, samples of type C (argon-plasma-treated sapphire) exhibit deviations from activated behavior, including downturns and bumps, together with much smaller power-law exponents extracted from the low-$T$ fits. The same samples also exhibit lower $Q_i$, indicating that the loss mechanisms captured by microwave measurements are accompanied by additional low-energy spectral weight in the superconducting state.

Within conventional superconducting electrodynamics, a non-exponential $\lambda(T)$ at $T\ll T_c$ generally reflects one (or a combination) of: (i) symmetry-imposed or accidental nodes, (ii) strong gap anisotropy in the presence of impurity scattering, or (iii) additional pair-breaking mechanisms that induce bound states in the gap. The first and second scenarios are not relevant to elemental Ta. However, significant differences in the low-$T$ response point to additional pair-breaking. In particular, these states can arise from two-level systems (TLS) and/or Yu-Shiba-Rusinov in-gap states \cite{Yu1965,Shiba1968,Rusinov1969,abogoda2025,he2025a}.

A natural candidate for this low-energy spectral weight is the same defect landscape that limits $Q_i$ in microwave resonators: the TLS and the structurally and chemically disordered layers that host them. Although losses are most commonly attributed to TLS (which may arise from different microscopic mechanisms), other pair-breaking defects and processes also influence the superconducting quasiparticle spectrum measured by a macroscopic probe, such as ours. Local disorder in surface oxide can modify bonding and the effective pairing environment, introduce resonant scattering channels, and potentially stabilize localized magnetic moments. Any of these effects can produce subgap states and broaden the coherence peaks, leading to an increased quasiparticle population that manifests as a measurable, non-activated, and sometimes irregular response, such as bumps and downturns in $\lambda(T)$ at low temperatures. This establishes a correlation between the structure of the metal/substrate layers (Fig.~\ref{fig:sample_info}) and our observations.

The markedly improved behavior of type-B films indicates that the Nb interlayer modifies the Ta/sapphire interface, thereby benefiting both crystallinity and defect chemistry. Several non-exclusive mechanisms are plausible. The Nb layer may act as a structural template that promotes higher-quality Ta nucleation at the interface, reduces the formation of a structurally disordered interfacial layer, and modifies oxygen kinetics through gettering and/or redistribution. In agreement with this picture, type-B samples show substantially improved microwave performance. The argon-plasma treatment used for type-C films may improve adhesion and influence the interfacial structure. Still, our superconducting-state $\chi(T)$ measurements indicate that residual low-energy excitations remain comparatively significant.

Finally, TDR-based spectroscopy probes the bulk-averaged electrodynamic response and therefore complements local or interface-specific techniques. A definitive microscopic assignment of the inferred low-energy states will require correlated measurements that directly target the interface region and possible magnetic degrees of freedom, such as low-temperature scanning tunneling spectroscopy, controlled oxide modification and passivation studies, electron spin resonance, and systematic dependence on film thickness and post-growth processing. Nevertheless, the present results demonstrate that precision penetration-depth measurements provide a sensitive and non-destructive screening tool for QIS material development.

\section{Summary}

In summary, we used precision measurements of the Meissner-state magnetic susceptibility — interpreted as superconducting gap spectroscopy — to compare Ta films intended for use in transmon qubit devices. Tantalum films directly deposited on $c$-axis sapphire substrates show signatures of localized low-energy excitations consistent with additional subgap spectral weight, and these films also exhibit the lowest quality factors. Films deposited on argon-plasma-treated substrates show only modest improvement in the susceptibility measurements. The best performance is found in films grown on a thin (5 nm) Nb overlayer deposited on sapphire, which also shows a clean, activated, low-$T$ response consistent with a fully gapped spectrum. It is important to note that different substrates result in notably different film morphologies and granular structures. Overall, non-destructive frequency-domain TDR spectroscopy provides a practical addition to the QIS materials characterization toolbox.

\begin{acknowledgments}
We thank James Sauls, Mehdi Zarea, Maria Iavarone, and John Zasadzinski for fruitful discussions. This work was supported primarily by the U.S. Department of Energy, Office of Science, National Quantum Information Science Research Centers, Superconducting Quantum Materials and Systems Center (SQMS), under Contract No. 89243024CSC000002. Fermilab is operated by Fermi Forward Discovery Group, LLC under Contract No. 89243024CSC000002 with the U.S. Department of Energy, Office of Science, Office of High Energy Physics. Ames Laboratory is supported by U.S. Department of Energy (DOE), Office of Science, Basic Energy Sciences, Materials Science $\&$ Engineering Division and is operated by Iowa State University for the U.S. DOE under contract DE-AC02-07CH11358. Thin films were grown at NIST. M.A.T. and K.R.J. were supported by DOE, BES, MSED at Ames National Laboratory.  
\end{acknowledgments}

%

\end{document}